\documentclass[11pt]{tCON2e}

\usepackage{amsmath}
\usepackage{amssymb}        
\usepackage{graphicx}

\usepackage{amsthm}
\usepackage{lscape}
\usepackage{arydshln}
\usepackage{slashbox}
\usepackage{subcaption}

\usepackage[
    pdftitle={Tool Eval},
    pdfauthor={ZL},
    pdfsubject={Tool Eval},
    pdfkeywords={},
    bookmarks=true,
    bookmarksnumbered=true,
    bookmarksopen=true,
    bookmarksopenlevel=2,
    colorlinks=true,
    pdfstartview=Fit,
    linktocpage=true,
    hyperindex=true,
    pdfpagemode=UseOutlines,    
    ]
{hyperref}

\hypersetup{
    bookmarksnumbered,
    pdfstartview={FitH},
    citecolor={blue},
    linkcolor={blue},
    urlcolor={blue},
    pdfpagemode={UseOutlines}
}

\PassOptionsToPackage{hyphens}{url}
\usepackage[hyphens]{url}
\usepackage{breakurl}

\usepackage{epstopdf}

\theoremstyle{plain}
\newtheorem{theorem}{Theorem}

\theoremstyle{definition}

\theoremstyle{remark}

\begin{document}


\articletype{Survey paper}

\title{\large A review and evaluation of numerical tools for fractional calculus and fractional order controls}

\author{
\large
\name{Zhuo Li\textsuperscript{a}
Lu Liu\textsuperscript{a}\textsuperscript{b}
Sina Dehghan\textsuperscript{a}
YangQuan Chen\textsuperscript{a}$^{\ast}$\thanks{$^\ast$Corresponding author. Mechatronics, Embedded Systems and Automation Laboratory (MESA Lab) \url{http://mechatronics.ucmerced.edu}. Email:\url{ychen53@ucmerced.edu}}
and Dingy\"{u}  Xue\textsuperscript{b}}
\affil{\textsuperscript{a}School of Engineering, University of California, Merced. 5200 N. Lake Road, Merced, CA 95343, USA
\textsuperscript{b}College of Information Science and Technology, Northeastern University, Shenyang 110819, China}
\received{June 2015}
}


\maketitle

\newpage

\begin{abstract}
\large
\begin{center}
\textbf{Abstract}
\end{center}
In recent years, as fractional calculus becomes more and more broadly used in research across different academic disciplines, there are increasing demands for the numerical tools for the computation of fractional integration/differentiation, and the simulation of fractional order systems. Time to time, being asked about which tool is suitable for a specific application, the authors decide to carry out this survey to present recapitulative information of the available tools in the literature, in hope of benefiting researchers with different academic backgrounds. With this motivation, the present article collects the scattered tools into a dashboard view, briefly introduces their usage and algorithms, evaluates the accuracy, compares the performance, and provides informative comments for selection.
\end{abstract}


\newpage

\section{Introduction}

The fractional calculus (FC) got birth 300 years ago, and the research on fractional calculus experienced its boom in the past decades \cite{ref:Mainardi_recent_FC, ref:Miller_Ross_intro_frac_book, ref:Agrawal_Machado_book_FC_advance}.
Besides the fundamental mathematical study, more and more researchers from different academic disciplines begin to utilize it in a varieties of subject-associated research, such as in biology and biomedical \cite{ref:Magin, ref:Bruce_Book_where}, sociology \cite{ref:Bruce_Book_social, ref:Book_of_Extremes}, economics \cite{ref:long_memo_new_model, ref:random_walk_wall_st_book}, and control engineering \cite{ref:Monje, ref:Yinchun_Automatica, ref:Zhuo_relay_id_fo}, etc.
Along with the rapid development of theoretical study, the numerical methods and practical implementation also made considerable progress \cite{ref:Machado_implementation, ref:Bohannan2, ref:Hartley6}.

Sharp tools are prerequisite to a successful job. In this paper, an extensive collection of Matlab based tools are exhibited for the numerical computation of fractional order (FO) integration/differentiation, as well as some toolboxes for engineering applications, with an emphasis on fractional order controls. A comprehensive table, table \ref{tb:fc_tools}, is created to list the recapitulative information of these tools in a dashboard view. Brief description and basic evaluation of these numerical algorithms are presented, in terms of usage, accuracy, unique features, advantages and drawbacks. Through such efforts, it is hoped that an informative guidance is provided to the readers when they face to the problem of selecting a numerical tool for a specific application.
Thanks to the authors of these tools. It is these pioneers who bring great convenience for the practical use of FC and FO control. While a text descriptive survey on some of the tools under discussion can be found in book \cite{ref:FO_sig_proc_2012}, and $28$ alternatives for the time-domain implementation of FO derivatives are documented in \cite{ref:Valerio_time_implem}, the present paper addresses more quantitative comparison and practical usage.

The rest of the paper are organized as follows: section \ref{sec:review} reviews $20$ selected numerical tools through brief description; section \ref{sec:evaluation} evaluates and compares the quantitative performance of the tools in three categories; section \ref{sec:com_for_sel} gives comments for tool selection based on empirical use.

\section{Review and description}
\label{sec:review}

This article mainly covers the tools for fundamental fractional calculus, such as the numerical computation of fractional integration/differentiation of a function or a signal, the Laplace transform of fractional differential equations \cite{ref:Igor}, etc. Since automatic control is one of the engineering disciplines that got the earliest exposure to fractional calculus \cite{ref:Bode_book, ref:Bagley, ref:Axtell, ref:Igor_FOPID_1999}, the tools for the application of fractional order controls are given more focus, associated with the authors' expertise.

\subsection{@fotf}
@fotf (fractional order transfer function) is a control toolbox for fractional order systems developed by Xue \emph{et al}. Most of the functions inside are extended from the Matlab built-in functions. In \cite{ref:Chen_tutorial}, the code and usage of the @fotf toolbox are described in very detail. It uses the overload programming technique to enable the related methods of the Matlab built-in functions to deal with FO models. The transfer function objects generated from it can be interactive with those generated from the Matlab transfer function class. Yet, the overloading of associated functions such as {\tt impulse()}, {\tt step()}, etc, lost the plotting functionality. As a work around, users can simply define a time vector as the second input to these functions. fotf toolbox supports time delay in the TF, e.g. {\tt fotf(a,na,b,nb,delay)}. It does not directly support transfer function matrix, hence, MIMO systems cannot be simulated directly. However, since it provides Simulink block encapsulation of the involved function {\tt fotf()}, multiple input/output relationship can be established by manually adding loop interactions in Simulink block diagrams. Therefore, the remark ``could'' is put in the ``MIMO'' column in table \ref{tb:fc_tools}, (where the `Delay' column denotes if the script/toolbox is able to handle time delay in the FO model; and the `MIMO' column denotes if the script/toolbox is able to handle MIMO FO models.).

A small drawback with @fotf is that the sampling time has relatively big impact on the accuracy, which has been remarked in the validation comments in \cite{ref:Chen_tutorial}.
Encouragingly, an update is upcoming according to the author.


\subsection{Ninteger}
Ninteger, non-integer control toolbox for Matlab, is a toolbox intended to help with developing fractional order controllers and assessing their performance, \cite{ref:Duarte3}. It uses integer order transfer functions to approximate the fractional order integrator/differentiator, $C(s) = k s^{\nu}, \ \nu \in \mathbb{R}$. It offers three frequency domain approximation methods,
\begin{enumerate}
    \item The CRONE methods, which uses a recursive distribution,
    \begin{eqnarray*}
        C(s) = k'\prod_{n=1}^N \frac{1+s/\omega_{zn}}{1+s/\omega_{pn}};
    \end{eqnarray*}
    \item The Carlson's method that solves $C^{\alpha}(s)$ using Newton's iterative method,
    \begin{eqnarray*}
        C_n(s) = C_{n-1}(s)\frac{(\alpha-1)C^{\alpha}_{n-1}(s)+(\alpha+1)g(s)}{(\alpha+1)C^{\alpha}_{n-1}(s)+(\alpha-1)g(s)};
    \end{eqnarray*}
    \item The Matsuda's methods, that approximates $C$ with a gain known at several frequencies.
    \begin{eqnarray*}
        C(s) &=& [d_0(\omega_0); \ \ \ (s-\omega_{k-1})/d_k(\omega_k)]^{+\infty}_{k-1}, \\
        d_0(\omega)&=& |C(j\omega)|, \ \ \ d_{k+1}(\omega) = \frac{\omega-\omega_k}{d_k(\omega) - d_k(\omega_k)}.
    \end{eqnarray*}
\end{enumerate}
It also provides Simulink block encapsulation of the involved functions, such as `nid' and `nipid' blocks. Moreover, it offers a user-friendly GUI for fractional order PID controller design.

There is a problem with ninteger toolbox in Matlab version $2013a$ or later. Without additional editing, it has conflicts with some built-in functions due to the overload editing of the Matlab built-in function {\tt isinteger()}. For example, calling the {\tt mean()} function will prompt an error.

\subsection{ooCroneToolbox}
The CRONE Toolbox, developed since the nineties by the CRONE team, is a Matlab and Simulink toolbox dedicated to applications of non integer derivatives in engineering and science \cite{ref:CRONE1}. It evolved from the original script version to the current object-oriented version. A good feature of the Crone toolbox is that some of the methods are implemented for MIMO fractional transfer functions. For example, executing {\tt sysMIMO=[sys,sys;sys2,sys2]} generates a two-input-two-output TF matrix. Many simulation results in the literature are obtained using the CRONE toolbox such as the design of centralized CRONE controller with the combination of the MIMO-QFT approach in \cite{ref:Oustaloup_MIMO}. Several other toolboxes are inspired by CRONE, e.g. ninteger and FOMCON.
A drawback of the CRONE toolbox is that time delay cannot be incorporated into the generated FO TF. Manually multiplying the delay to the {\tt frac\_tf} object does not work either because the {\tt exp()} operation is not overloaded by {\tt frac\_tf} class. CRONE is a toolbox much more powerful than merely simulating fractional order systems. In spite of this basic functionality, it is also capable of fractional order system identification and robust control analysis and design.

\subsection{FOMCON}
\begin{figure}[h]
\centering
\includegraphics[width=0.35\textwidth]{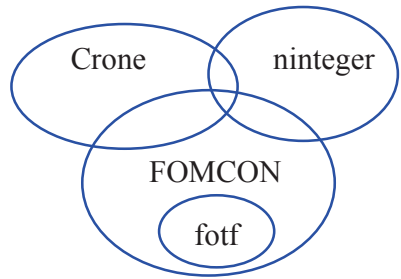}
\caption{FOMCON's relation to other numerical tools, \cite{ref:FO_ID_MS_Thesis}.}
\label{fig:FOMCON_relation}
\end{figure}

The FOMCON (Fractional-Order Modeling and Control) toolbox is developed by Tepljakov \emph{et. al}, \cite{ref:FOMCON}. Its kernel utilizes the algorithms in FOTF, Ninteger and Crone. It encapsulates some of the major functionalities of those three toolboxes, and builds a GUI shell on top, aiming at extending classical control schemes for FO controller designs. The relation of FOMCON with the three toolboxes is shown in figure \ref{fig:FOMCON_relation}. Some notable changes/patches to the original FOTF are:
\vspace{-0.1cm}
\begin{itemize}
    \item[\tiny$\bullet$] {\ttfamily newfotf()} uses the string parser to enable users to input TF as a string;
    \item[\tiny$\bullet$] {\tt tf2ss()} is overloaded and {\tt foss()} is added, which makes the conversion between an FO TF object and an FO state space object possible. The CRONE toolbox is also able to do the task, yet the script is encrypted in Matlab P code format.
\end{itemize}

\subsection{M-L functions}
M-L functions, as the name implies, are Matlab functions developed for numerically computing the Mittag-Leffler function (definitions can be found in \cite{ref:Magin} etc). There are several versions of code by different authors available in the literature. Five of them are listed in table \ref{tb:fc_tools}, where
\begin{enumerate}
    \item {\tt mlf($\alpha, \beta, x, p$)} is for the calculation of the 2-parameter M-L function in the form of $E_{\alpha,\beta}(x)$ with the precision of $p$ for each element in $x$;
    \item {\tt ml\_func([$\alpha, \beta, \gamma, q], z, n, \varepsilon_0$)} is capable of computing the M-L function with either 1, 2, 3, or 4 parameters, and the script is available in the books \cite{ref:Xuedingyu} or \cite{ref:Monje}. It uses the fast truncation algorithm to improve the efficiency, and embeds the {\tt mlf()} in the file such that when the fast truncation algorithm is not convergent, solution is guaranteed by trading off some efficiency;
    \item {\tt ml\_fun($\alpha, \beta, x, n, \varepsilon_0$)} ($\alpha>0, \beta>0$) is also for 2-parameter M-L function with error tolerance of $\varepsilon_0$, which is implemented using C-MEX .dll (dynamic-link library) technique and can be used in Simulink through s-functions;
    \item {\tt gml\_fun($\alpha, \beta, \gamma, x, \varepsilon_0$)} calculates the generalized M-L function with 3 parameters in the form of $E_{\alpha,\beta}^{\gamma}(x)$, \cite{ref:generalized_ML};
    \item {\tt ml($x, \alpha, \beta, \gamma$)} can calculate the M-L function with either 1, 2, ro 3 parameters.
\end{enumerate}

Alternatively, the generalized hypergeometric function {\tt [pfq]=genHyper (a,b,z, lnpfq, ix, nsigfig)} in \cite{ref:genHyper}, or {\tt [y,tt,nterms]=pfq (a,b,z,d)} in \cite{ref:pFq} can also achieve the numerical computation of the generalized M-L functions under certain conditions. For more details, refer to \cite{ref:generalized_ML_Chaurasia}.

\subsection{NILT}

The inversion of Laplace transform is fundamentally important in the applications of Laplace transform method. It can be carried out with one of the following three approaches: 1). analytical solution using definition and basic properties; 2). Laplace transform tables; and 3). numerical computation. While analytical solutions are usually too hard to be obtained, and tables do not cover arbitrary cases, the numerical computation becomes an inevitable way. Among the numerous algorithms for numerical inversion of Laplace transform (NILT), {\tt NILT} in \cite{ref:INVLAP_1982, ref:Liangjinsong_phd_thesis} and the ``improved NILT'' in \cite{ref:Lubomir_NILT1, ref:Lubomir_NILT_improved, ref:Lubomir_NILT} have relatively bigger literature exposure.
Lubomir's NILT method applies the fast Fourier Transformation (FFT) and the $\varepsilon$-algorithm to speed up the convergence of infinite complex Fourier series. A very detailed description and performance evaluation of these methods is available in \cite{ref:Shenghu_NILT}. Hence, repetitive comparison among different NILTs are not presented here. Focus is mainly put on the comparison between NILT and other numerical methods.

\begin{figure}[h]
\centering
\includegraphics[width=0.75\textwidth]{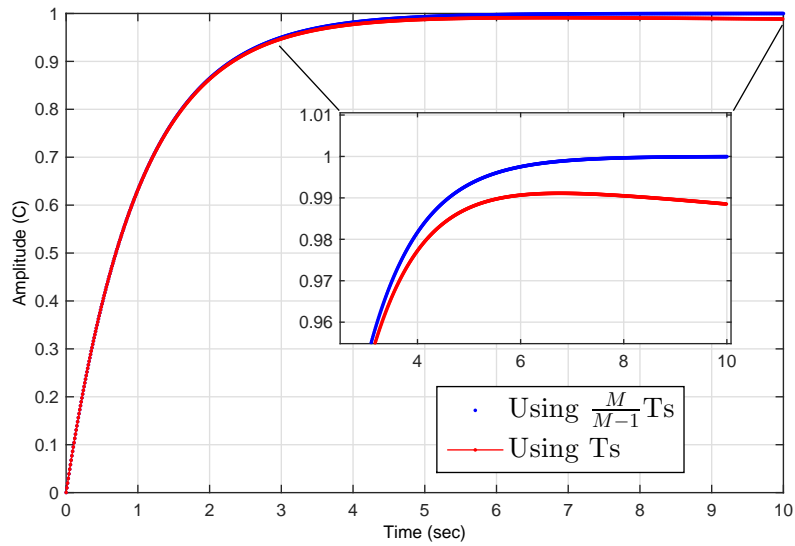}
\caption{Computation error of NILT caused by mis-assignment of Ts.}
\label{fig:NILT_ts_issue}
\end{figure}

A good feature of the two NILT code is that both support the direct input of time delay in the form of {\tt exp(-Ls)}. Yet, {\tt INVLAP()} gives some glitch at the end of the delay, for example, {\tt [x,y]=INVLAP('1/(s* (s\string^0.5+1)) * exp(-s)', 0.01,10,1000)}.
There is a tricky part need to be noted in evaluating the computational error of {\tt NILT}. If the same initial, terminating and sampling time ($t_0, t_f\ \rm{and} \ T_s$) for other tools are used in the script, the {\tt NILT} actually computes one point less than the other tools which use regularly spaced time vector. That is because: let $M = \frac{t_f-t_0}{T_s}$ represent the amount of points computed by {\tt NILT}, then, the time interval is actually $T_s' = \frac{t_f-t_0}{M-1}$ due to the script {\tt t=linspace(0,tm,M)}. Whereas the conventional assignment of time vector ({\tt t=t0:Ts:tf}) generates {\tt M+1} points. In order to compute the same amount of points aligned to the time stamps used for baseline analytical solution, the time vector for analytical computation needs to be adjusted so as to adapt to that used by {\tt NILT}. This means to let analytical computation use the time vector generated by {\tt NILT}, which can be achieved by either 1). {\tt t=0:M*Ts/(M-1):M*Ts}, or 2). {\tt t=linspace(0,tm,M)}. This cannot be done the other way around, i.e. replaced by {\tt t = 0:Ts:M*Ts-Ts} nor {\tt t=linspace(0,tm-Ts,M)}. Otherwise, cumulated computation error will cause inaccuracy of the final simulation result. Alternatively, if $t_f$ is not a concern, user can assign one point less to {\tt M} in the {\tt NILT} script while keeping $T_s$ unchanged. Thus, {\tt NILT} generates the same time stamps except a $t_f$ shortened by one sampling period. The difference in dealing with time vectors can be easily visualized if longer sampling time is assigned. An example of the resulting computation error is demonstrated in figure \ref{fig:NILT_ts_issue}. Similar time stamp assignment issue exists in {\tt INVLAP()}. In addition, the initial time stamp is not allowed to be $0$ due to the constraint in the {\tt INVLAP()} script.

\subsection{dfod}
DFOD (Digital Fractional Order Differentiator/integrator) is a set of Maltab functions writhen by Petr\'{a}\v{s}, for the approximation of fractional order differentiators and integrators.
There are three versions of dfod:
\begin{enumerate}
    \item {\tt dfod1()} is the infinite impulse response (IIR) type based on continued fraction expansion (CEF), shown in equation (\ref{eqn:CFE_approx}), of weighted operator with the mixed scheme of the trapezoidal (Tustin) rule and the backward difference (Euler) rule, \cite{ref:Petras_dfod1};
        \begin{eqnarray}
        \label{eqn:CFE_approx}
        Z\{D^{\alpha} x(t)\} = CFE\{(\frac{1-z^{-1}}{T})^{\alpha}\}X(z) \approx (\frac{1}{T})^{\alpha}\frac{P_p(z^{-1})}{Q_q(z^{-1})} X(z).
        \end{eqnarray}
    \item {\tt dfod2()} is the finite impulse response (FIR) type based on power series expansion (PSE), shown in equation (\ref{eqn:PSE_approx}), of the backward difference (Euler) rule, \cite{ref:Petras_dfod2};
        \begin{eqnarray}
        \label{eqn:PSE_approx}
        {D^{ \mp \alpha }}(z) = \frac{1}{{{{(1 - {z^{ - 1}})}^{ \pm \alpha }}}} = \frac{{{T^{ \mp \alpha }}}}{{\sum\limits_{j = 0}^\infty  {{{( - 1)}^j}\left(\!\!\! {\begin{array}{*{20}{c}}
        { \pm \alpha }\\
        j
        \end{array}}\!\!\! \right){z^{ - j}}} }} \approx \frac{{{T^{ \mp \alpha }}}}{{{Q_q}({z^{ - 1}})}}
        \end{eqnarray}
    \item {\tt dfod3()} is a new IIR type based on power series expansion of the trapezoidal (Tustin) rule, \cite{ref:Petras_dfod3}.
    \begin{eqnarray}
    \rm{Euler:} \ s^{\alpha} \approx \left[\frac{1-z^{-1}}{T}\right]^{\alpha}, \ \  \ \ \rm{Tustin:} \ s^{\alpha} \approx \left[\frac{2}{T}\frac{1-z^{-1}}{1+z^{-1}}\right]^{\alpha}.
    \end{eqnarray}
\end{enumerate}

\noindent
There are other FO algorithms based on IIR, such as {\tt newfod()} by Chen, \cite{ref:Chen_IIR}.

Regarding discretization, besides the aforementioned methods used in the various tools, other methods exist such as the Prony's technique, direct discretization, the binomial expansion of the backward difference, etc, \cite{ref:book_FO_Riccardo}.

\subsection{IRID}
The impulse response invariant discretization (IRID) is a family of functions designed by Chen, Li, Sheng \emph{et al.} \cite{ref:Liyan_irid, ref:Chen_irid_fod}, for different approximation purposes based on the algorithm as its name implies. It includes the following members:
\begin{enumerate}
    \item {\tt irid\_fod()} is designed to compute a discrete-time finite dimensional ($z$) transfer function to approximate a continuous irrational transfer
        function $s^\alpha$ where `$s$' is the Laplace transform variable and $-1<\alpha<1$. It has been tested that the algorithm still works for $\alpha>1$ and $\alpha<-1$, by removing the input checking statement.
    \item {\tt irid\_doi()} is for the approximation of distributed order integrator $\int _a^b \frac{1}{s^{\alpha}} d \alpha $, where `$a$' and `$b$' are arbitrary real numbers in the range of (0.5 ,1), and $a<b$.
    \item {\tt irid\_dolp()} is for the approximation of a continuous-time fractional order low-pass filter in the form of $1/(\tau s +1)^{\alpha}$
    \item {\tt irid\_fsof()} is for the approximation of fractional second order filter in the form of $ 1/(s^2 + a s + b)^{\alpha}$ where $0<\alpha<1$.
    \item {\tt BICO\_irid()} is for the approximation of BICO (Bode's Ideal Cut-Off) transfer function in the form of $  1/(s/w_0+ \sqrt{(s/w_0)^2+1})^\alpha $, where $\alpha>0$.
\end{enumerate}

\subsection{ora\_foc}
{\tt ora\_foc()} is for the approximation of fractional order differentiators, $\frac{1}{s^\alpha}$, \cite{ref:Book_Xue_Chen_linear_feedback}, using the Oustaloup-Recursive-Approximation method described in \cite{ref:Oustaloup_ora}.

\subsection{fderiv}
{\tt fderiv()} calculates the fractional derivative of order $\alpha$ for the given function $r(t)$ using the Gr\"{u}nwald-Letnikov (G-L) definition, \cite{ref:Fractional_differentiator}. The input of the given function is represented by a vector of signal values.
There is an improved implementation of this function, {\tt fgl\_deriv()}, by Jonathan, which uses vectorization for faster computation with Matlab, \cite{ref:Fractional_derivative}.

\subsection{glfdiff}
{\tt glfdiff(y,t,$\alpha$)} (G-L finite diff) is a Matlab function written by Xue \emph{et al.} \cite{ref:Xuedingyu_book} for calculating the $\alpha^{th}$ derivative of a given function, whose inputs $y,\ t$ are the signal and time vectors. It is based on the forward finite difference approximation of the G-L definition,
\begin{eqnarray}
{_a}D^{\alpha}_t f(t) \approx \frac{1}{h^{\alpha}} \sum_{j=0}^{(t-1)/h} \omega_j^{(\alpha)}f(t-jh),
\end{eqnarray}
where the binomial coefficients are recursively calculated, \cite{ref:Xuedingyu_book}:
\begin{eqnarray}
\omega_0^{(\alpha)} = 1, \ \ \ \omega_j^{(\alpha)} = \left(1- \frac{\alpha+1}{j} \right) \omega_{(j-1)}^{(\alpha)}, \ \ \ j=1,2,\ldots
\end{eqnarray}

\subsection{Fractional differentiation and integration}
Many of the above functions approximate the fractional order integral or derivative operator. This Matlab function calculates the $\alpha^{th}$ order derivative or integral of a function, defined in a given range through Fourier series expansion. The necessary integrations are performed with the Gauss-Legendre quadrature rule, \cite{ref:FO_diff_and_int}. Three examples are offered in this package, namely FO differ/integral of identity, cubic polynomial and tabular functions, respectively. The main call function is {\tt fourier\_diffint()}.

\subsection{FIT}
FIT is the Fractional Integration Toolbox developed by Santamaria Laboratory at the University of Texas at San Antonio, \cite{ref:FIT_download}. It is for the numerical computation of fractional integration and differentiation of the Riemann-Liouville (R-L) type, and is designed for large data size, which allows parallel computing of multiple fractional integration/differentiation on GPUs (graphical processing units). The extrapolation and interpolation algorithms used by this toolbox are implemented in C++ and are integrated with Matlab via MEX mechanism. Detailed explanation can be found in \cite{ref:FIT}.

\subsection{DFOC}
DFOC, written by Petr\'{a}\v{s}, is a digital version of the Fractional-Order PID Controller of the form:
\begin{eqnarray}
C(s) = K + T_i\frac{1}{s^m} + T_d s^d.
\end{eqnarray}
It provides a transfer function of the FO PID controller for given parameters, \cite{ref:Petras_DFOC}.

\subsection{FOPID}
The fractional order PID (FOPID) controller toolbox, presented by Lachhab \emph{et al.}, is for the design of robust fractional order $PI^\alpha D^\beta$ controllers, \cite{ref:Nabil_FOPID}. The tuning rules for the parameters follow those promoted in \cite{ref:Luoying1} and \cite{ref:Luoying}. Thus, the fractional order PID tuning is converted to a 5-parameter optimization problem. This toolbox utilize the ``non-smooth'' $H_{\infty}$ synthesis in \cite{ref:NonsmoothHinfinity} to perform the minimization. For now, there is not a publicly available source for download.

\subsection{Sysquake FO PID}
In \cite{ref:Sysquake}, Pisoni \emph{et al.} presented an interactive tool for fractional order PID controllers developed on the Sysquake software environment, which is a similar effort with that for integer order PIDs done by {\AA}str\"{o}m \emph{et al.} in \cite{ref:Sysquake_Astrom}. Sysquake is a numerical computing environment based on a programming language mostly-compatible with Matlab. However, the interactive tool for FO PID runs in the Sysquake environment instead of Matlab. Hence, it is not reviewed in detail here.

\subsection{FOCP}
In \cite{ref:FOCP_Tricaud_paper}, Tricaud and Chen \emph{et al.} formulated the Fractional Optimal Control Problems (FOCP) into the integer order format by using a rational
approximation of the fractional derivative obtained from the singular value decomposition (SVD) of the Hankel matrix of the impulse response. Then, RIOTS\_95 \cite{ref:Zhuo_biao, ref:RIOTS_Tricaud_IFAC} is used to perform the optimization. The scheme is potentially able to solve any type of FOCPs and is implemented in Matlab for public accessibility, \cite{ref:FOCP_Tricaud}. It supports MIMO FO optimal control, but does not handle time delay due to the limitation of RIOTS.

\subsection{FSST}
\begin{figure}[h]
\centering
\includegraphics[width=0.5\textwidth]{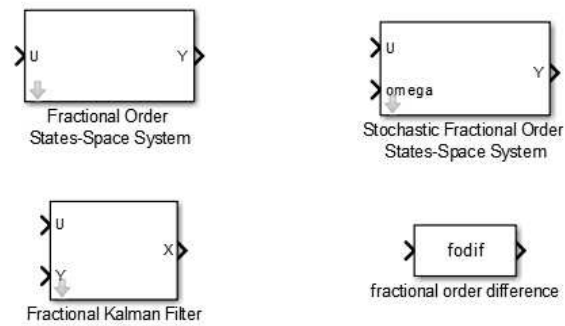}
\caption{The Simulink block set provided in FSST.}
\label{fig:FSST_Simulink}
\end{figure}

\begin{figure}[h]
\centering
\includegraphics[width=0.75\textwidth]{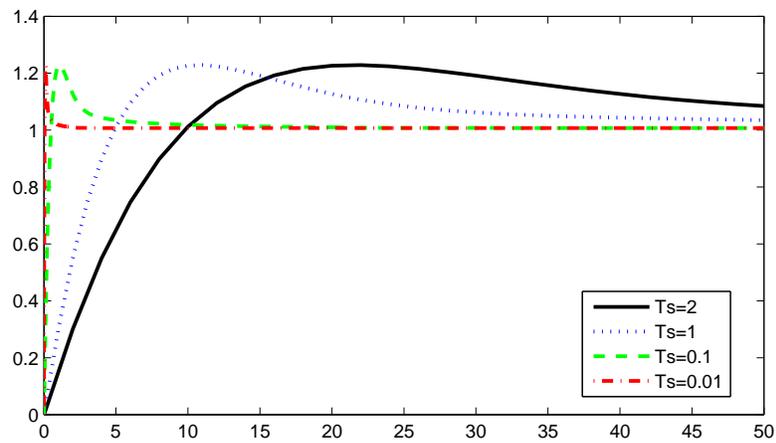}
\caption{The impact of simulation step size on the FSST toolbox.}
\label{fig:Ts_imp_FSST}
\end{figure}

FSST is a simulation toolkit in Matlab/Simulink for the fractional order discrete state-space system education. The toolkit consists of a set of C-MEX s-functions which are encapsulated in Simulink blocks. Several typical fractional order system simulation examples are provided as shown in figure \ref{fig:FSST_Simulink}, such as the fractional order state-space model and the fractional Kalman filter (FKF), \cite{ref:fsst_Dominik_IFAC}. The version $1.7$ is available for free download at \cite{ref:fsst_1_7}.
Two of the superior strengths of FSST are: 1.) it can directly simulate MIMO systems since it is a Simulink block kit handling state space representations; 2.) it is able to incorporate the initial conditions into the dynamic equations to be simulated, which is a unique feature among all the aforementioned tools. The drawback of FSST is that the step size has large impact on the simulation results, even larger than the impact by ``cilcular'' buffer size. A sample illustration is plotted in figure \ref{fig:Ts_imp_FSST}.

\subsection{Fractional variable orders}
All the above tools/toolboxes (except {\tt irid\_doi()}) deal with constant fractional orders. Yet, there exists a type of differentiations that have fractional variable orders (FVO). The definitions in the G-L format are given as follows, \cite{ref:FVO_Lorenzo}:
\begin{theorem}[The $1^{st}$ type FVO]
\begin{eqnarray}
{_0}D_t^{\alpha(t)}f(t) = \lim_{h\rightarrow 0} \frac{1}{h^{\alpha (t)}} \sum _{r=0}^{n} (-1)^r \left(\!\!\! {\begin{array}{*{20}{c}}
        { \alpha (t) }\\
        r
        \end{array}}\!\!\! \right)
        f(t - rh).
\end{eqnarray}
\end{theorem}
\noindent
The $2^{nd}$ and $3^{rd}$ types can be found in the same reference.

Regarding the fractional variable order differentiation, there are dedicated tools.
Podlubny \emph{et al.} offers a matrix approach that unifies the numerical differentiation of integer order and the n-fold integration, using the so-called triangular strip matrices,
\cite{ref:Igor_matrix_approach}. It is available for download at \cite{ref:Igor_matrix_approach_download} and can be applied on the solution to FODEs and FPDEs.

Sierociuk \emph{et al.} provides a C-MEX s-function based Simulink toolkit, ``fvoderiv'', for this purpose, \cite{ref:fvoderiv_Dominik}. It supports real-time-workshop.

The toolbox ``vod'' created by Val\'{e}rio \emph{et al.} calculates variable fractional or complex order derivatives. R-L, Caputo and G-L definitions are provided; the three types of definitions in \cite{ref:FVO_Lorenzo} are all considered. Fuzzy supervised implementations in Simulink are also provided, \cite{ref:FVO_Valerio}.

\subsection{FO root locus}
\begin{figure}[h!]
\centering
    \begin{subfigure}[b]{0.49\textwidth}
        \includegraphics[width=\textwidth]{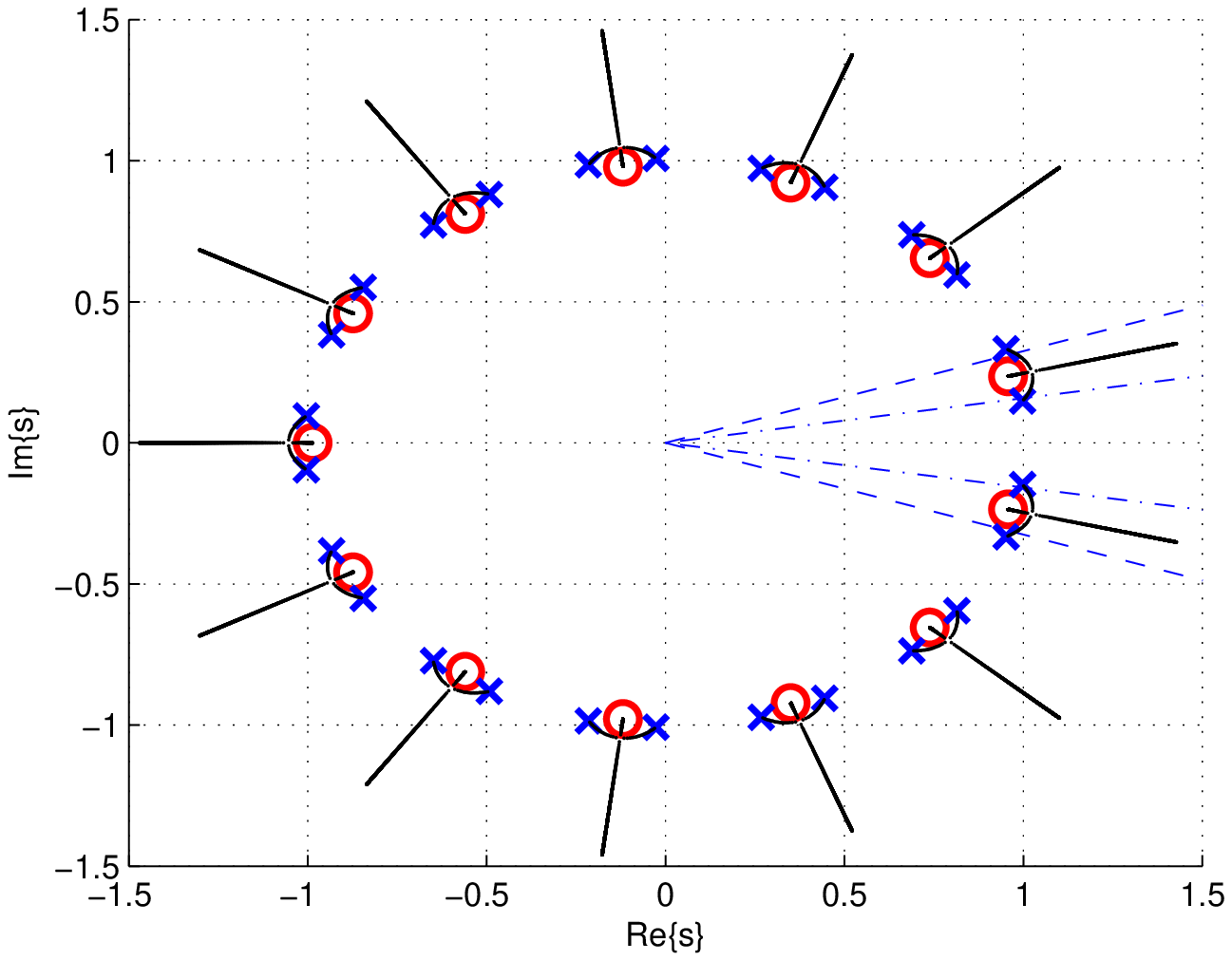}
        \caption{On $s$-plane.}
        \label{fig:eg2_3_3_a}
    \end{subfigure}
    \begin{subfigure}[b]{0.49\textwidth}
        \includegraphics[width=\textwidth]{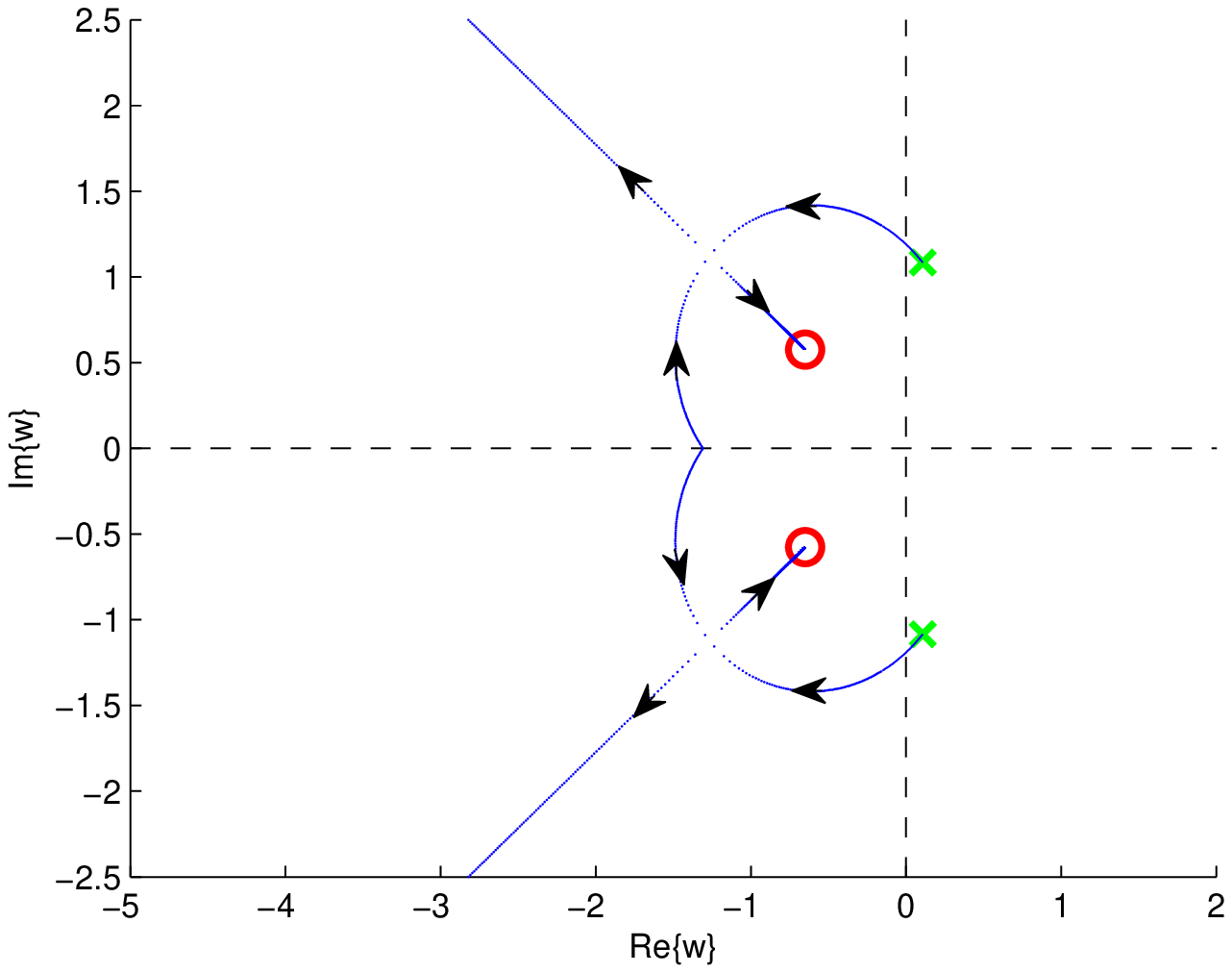}
        \caption{On $w$-plane.}
        \label{fig:eg2_3_3_b}
    \end{subfigure}%
    \caption{The RL plot of equation (\ref{eqn:eg_2.3.3}) on different planes.}
    \label{fig:eg2_3_3}
\end{figure}

Three Matlab based scripts for plotting root locus of fractional order TFs are available. Two early works are {\tt frlocus()} in \cite{ref:flocus_matlab}, and the code attached in the paper \cite{ref:rlocus_Machado_with_code} by Machado \emph{et al.}. The other is {\tt forlocus()} developed by the author which is listed in the last row in table \ref{tb:fc_tools} and can be downloaded from \cite{ref:FO_RL_mine}. Besides, the newest version of @fotf toolbox also features the root locus plot of FO systems. Figure \ref{fig:eg2_3_3} shows a demonstrating plot of the root locus of the following fractional order transfer function,
\begin{eqnarray}
\label{eqn:eg_2.3.3}
G(s) = \frac{1.2s^{1.3}+1}{0.8s^{2.6}+0.6s^{1.3}+1},
\end{eqnarray}
where figure \ref{fig:eg2_3_3_a} shows the plot on the Laplace s-plane and figure \ref{fig:eg2_3_3_b} shows the plot on the $w=s^{1.3}$ plane.
A closer view of the $2^{nd}$ quadrant in figure \ref{fig:eg2_3_3_b} tells that the root locus in this example has two branches on the first layer of the Riemann sheet \cite{ref:RiemannSurfaces, ref:RiemannSurfacesBook}. One starts from the pole marked in green, and the other is from the next Riemann sheet. As the system gain varies, they aggregate at $(-1.25+1.1i)$ and then bifurcate. One approaches to the open loop zero marked in red and the other goes to infinity.

\subsection{Other tools}
Text description of a few tools listed above can also be found in \cite{ref:FO_in_Matlab_Petras}.
There are other fractional calculus related tools or Matlab scripts available for specific applications, such as the fractional Fourier transform (FrFT) \cite{ref:FrFT_book_2001, ref:FrFT_paper_2003}, closed-form solutions to linear fractional order differential equations, {\tt fode\_sol()} \cite{ref:Monje}, the M-L random number generator {\tt mlrnd()}  \cite{ref:mlrnd}, digital fractional order Savitzky-Golay differentiator \cite{ref:Dali_Golay_filter}, and the functions for simulating fractional-order chaotic systems \cite{ref:Petras2}, etc. Considering the scope of research, they are not enumerated here and only fundamental FC and FO control related tools are reviewed.


\section{Evaluation and comparison}
\label{sec:evaluation}
\subsection{Comparison I}
To evaluate the collected tools, several groups of benchmark problems and inputs are designed. For the FO control toolboxes, the following problems are used,
\begin{enumerate}
    \item Baseline model: first order transfer function,  $$ g_{b}(s) = \frac{1}{s+1},$$
    whose time domain analytical solution of its step response is: $y(t) = 1- e^{-t}$;
    \item Impulse response of half order integrator: $$ g_{hint}(s) = \frac{1}{\sqrt{s}},$$
    whose time domain analytical solution is: $\frac{1}{\sqrt{\pi t}}$;
    \item TF with a half order pole: $$ g_{hp}(s) = \frac{1}{\sqrt{s}+1},$$
    whose time domain analytical solution is: $$\frac{1}{\sqrt{t}}E_{\frac{1}{2},\frac{1}{2}}(-\sqrt{t}), \ \rm{or \ equivalently,} \ \frac{1}{\sqrt{\pi t}}-e^{t} erfc [\sqrt{t}];$$
    \item The commensurate order TF: $$ g_{com}(s) = \frac{6s^{1.2}+s^{0.8}+2s^{0.4}+3}{5s^{1.6}+s^{0.8}+2};$$
    \item Step response of the irrational order TF: $$ g_{ir}(s) = \frac{2s^{\sqrt{3}}+1}{s^{\sqrt{5}}+3s^{\sqrt{2}}+1}.$$
\end{enumerate}

\begin{figure}[h]
\centering
\includegraphics[width=0.75\textwidth]{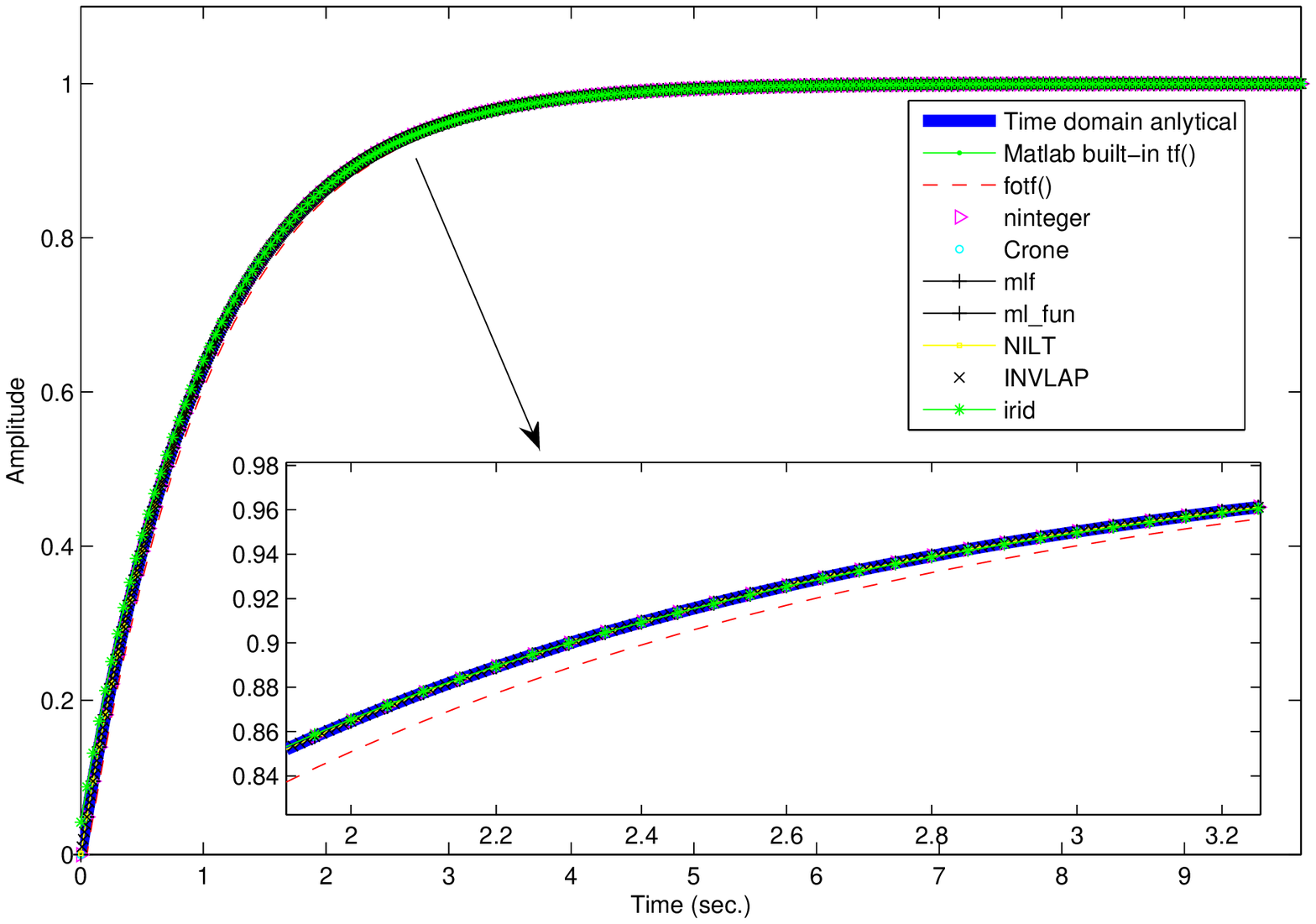}
\caption{Comparison of the step responses of problems 1.}
\label{fig:first_order_step}
\end{figure}

\begin{figure}[h]
\centering
\includegraphics[width=0.75\textwidth]{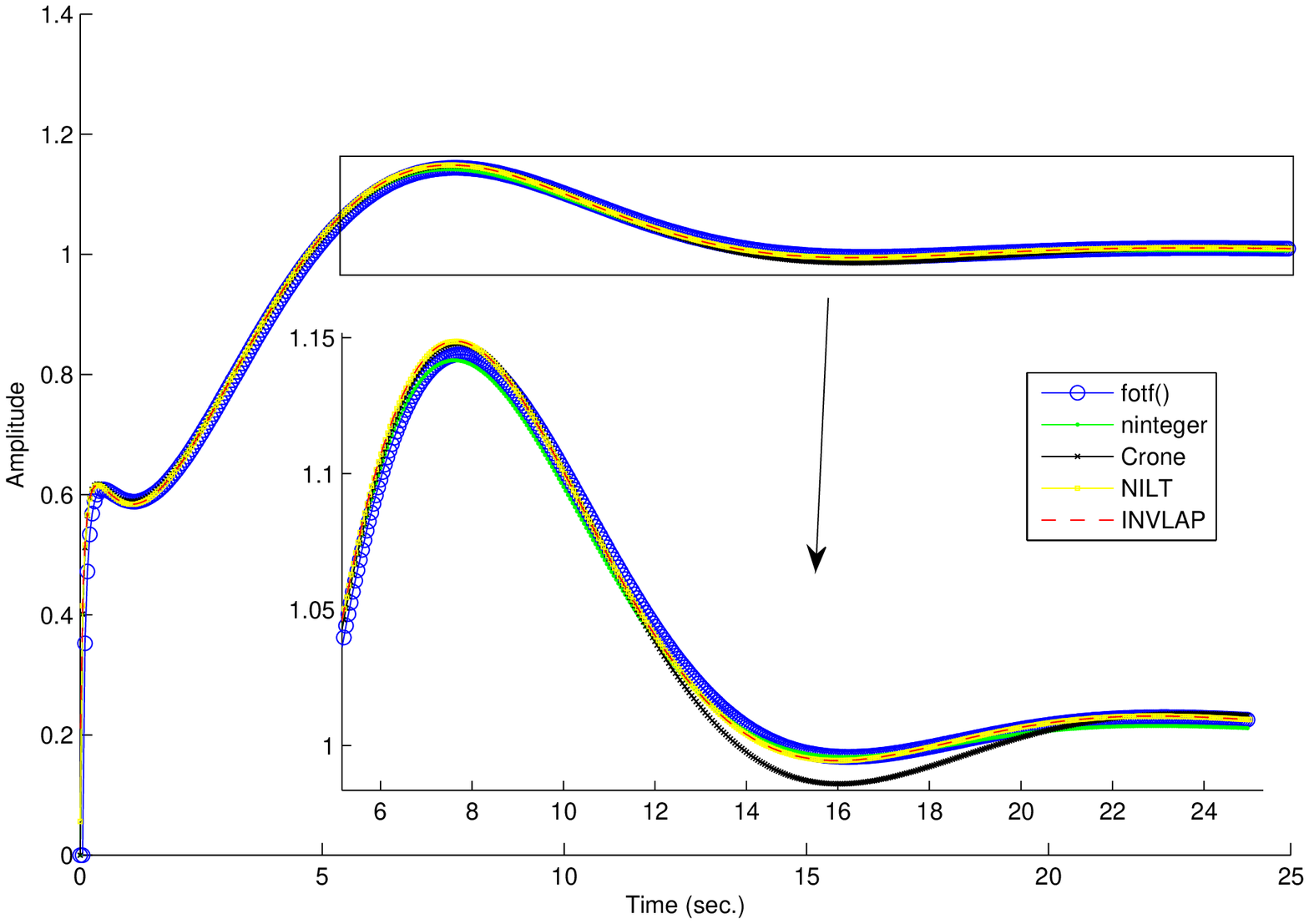}
\caption{Comparison of the step responses of problem 5.}
\label{fig:irrational_order_step}
\end{figure}

The accuracy is quantified by the conventional integral absolute error (IAE) criteria, $S = \int_{0}^{T}|e(t)|dt$. All comparison have been kept as fair as possible. The numerical values of the time domain analytical solution using Matlab built-in functions are assumed to be accurate and is adopted as the baseline. The computational errors when $T_s=0.05$ are summarized in table \ref{tb:tools_eval_error},
where the row indices represent the methods numbered in table \ref{tb:fc_tools}, and the column indices represent the test problems respectively. Besides, `M' denotes the Matlab built-in TF and `-' means the underlying method is not applicable for the test problem. Two sample plots of the step responses of problems $1$ and $5$ are shown in figures \ref{fig:first_order_step} and \ref{fig:irrational_order_step}. For problems $4$ and $5$, since analytical solution is hard to obtain, all methods are compared to the values computed by fotf.

\begin{figure}[h]
\centering
    \begin{subfigure}[b]{0.75\textwidth}
        \includegraphics[width=\textwidth]{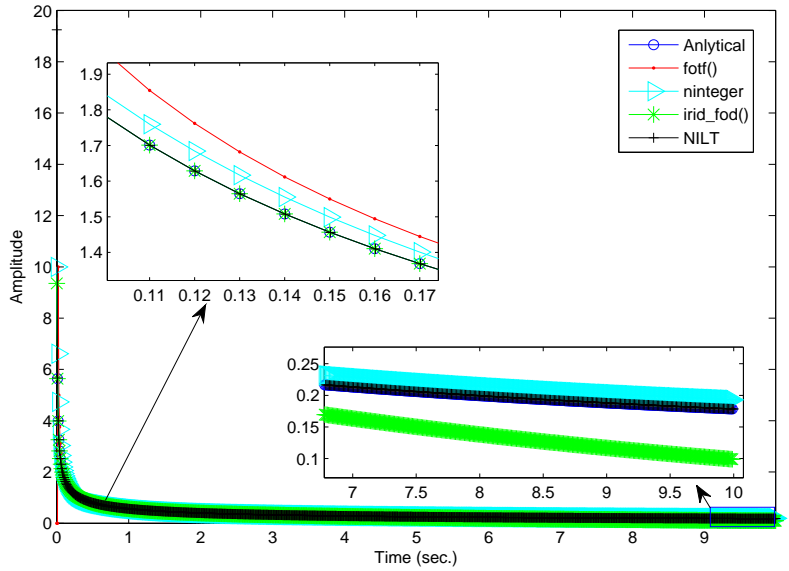}
        \caption{{\tt Ts=0.01}}
        \label{fig:half_order_impulse_Ts_0.01}
    \end{subfigure}\\%
    ~~ 
    \begin{subfigure}[b]{0.75\textwidth}
        \includegraphics[width=\textwidth]{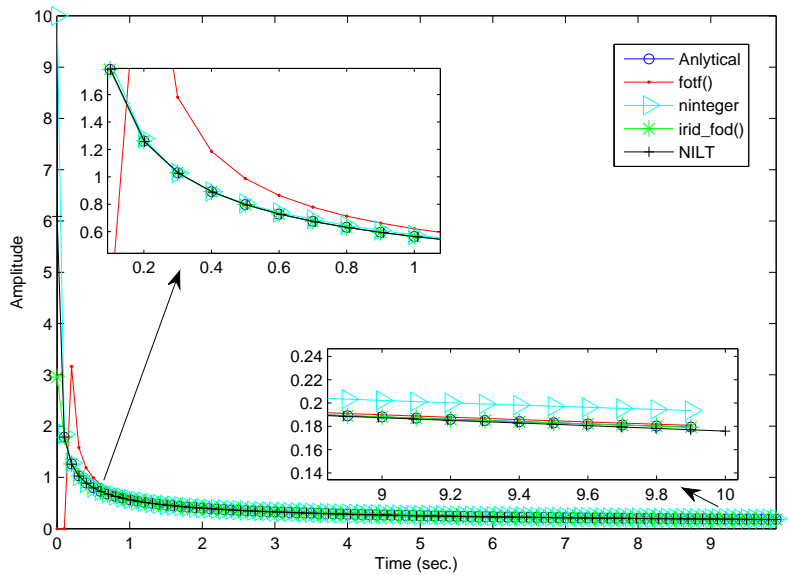}
        \caption{{\tt Ts=0.1}}
        \label{fig:half_order_impulse_Ts_0.1}
    \end{subfigure}%
    \caption{Comparison of the impulse responses of the half order integrator.}
    \label{fig:half_order_impulse}
\end{figure}

For the impulse response of the half-order integrator, the first point is ignored for error calculating because it is infinity. Two graphic views of the comparison are shown in figures \ref{fig:half_order_impulse_Ts_0.01} and \ref{fig:half_order_impulse_Ts_0.1}, with {\tt Ts=0.01} and {\tt Ts=0.1} respectively.

\begin{figure}[h]
\centering
\includegraphics[width=0.75\textwidth]{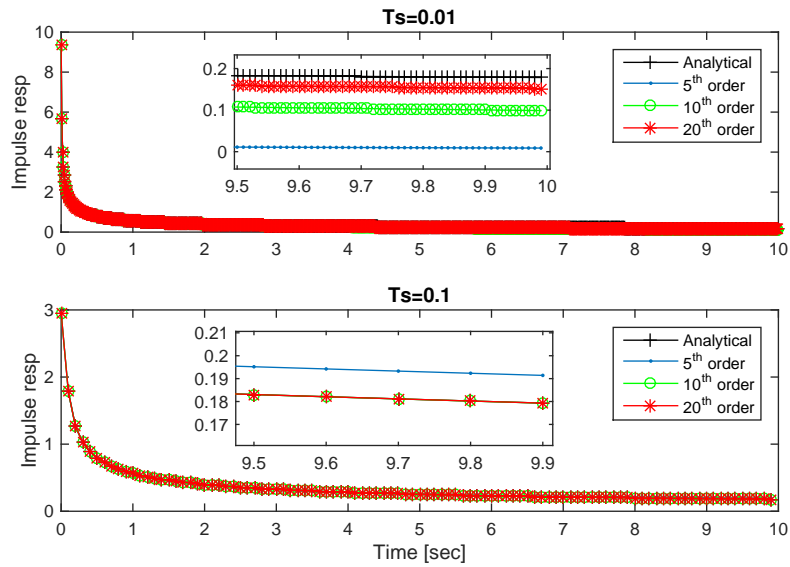}
\caption{The $T_s$ and order impact on {\tt irid\_fod()}.}
\label{fig:irid_Ts_order_relation}
\end{figure}
\begin{figure}[h]
\centering
\includegraphics[width=0.75\textwidth]{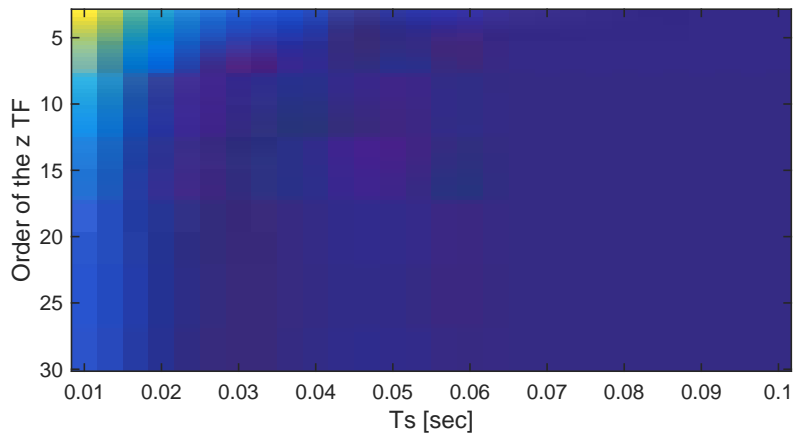}
\caption[Error heat map of {\tt irid\_fod()}]{The heat map of approximation error of {\tt irid\_fod()} versus the order and sampling time.}
\label{fig:irid_error_hMap}
\end{figure}

As stated in \cite{ref:Chen_irid_fod}, {\tt irid\_fod()} uses finite dimensional (z) TF as the approximation method. Hence, the order of the (z) TF has impact on the approximation accuracy. The error listed in table \ref{tb:tools_eval_error} is based on the $10^{th}$ order approximation. An illustrative plot is shown in figure \ref{fig:irid_Ts_order_relation}. The sampling time also has impact on its accuracy. The anti-intuitive fact is that relatively greater {\tt Ts} gives higher accuracy. A heat map of the error on the field of {\tt Ts=0.01:0.001:0.1} and {\tt order=3:30} is plotted in figure \ref{fig:irid_error_hMap}. At some particular high orders, ``rank deficient'' would occur during the call of {\tt prony()} inside {\tt irid\_fod()}. Users can choose appropriate orders according to their specific accuracy requirement.


The analytical expression of M-L function is a summation of infinite terms. Hence, it is not surprising to see the numerical computation induced error in the results.


\subsection{Comparison II}

{\tt fderiv()}, {\tt glfdiff()}, {\tt fourier\_diffint()} and FIT are integration /differentiation tools for functions. For this group of tools, the following two problems are designed to compare the performance.
\begin{enumerate}
    \item Half order derivative of the function $y(t) = 3t$ on the interval of $[0,5]$, whose analytical solution is,
    \begin{eqnarray}
    {_0}D_t^{0.5}y = \frac{3\Gamma(2)}{\Gamma(1.5)}\sqrt{t}.
    \end{eqnarray}
    \item $0.75$ order integration of the function $y(t) = \sqrt{t}$, whose analytical solution is,
    \begin{eqnarray}
    {_0}D_t^{-0.75}y = \frac{\Gamma(1.5)}{\Gamma(2.25)}t^{1.25}.
    \end{eqnarray}
\end{enumerate}

\begin{figure}[h]
\centering
\includegraphics[width=0.75\textwidth]{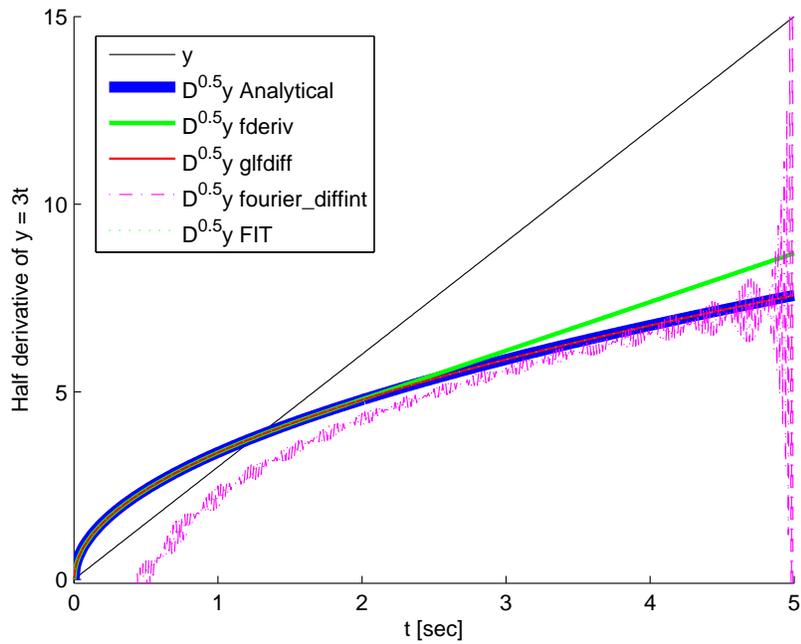}
\caption[Comparison of half derivative of function $y(t) = 3t$]{Comparison of half derivative of function $y(t) = 3t$, using {\tt fderiv()}, {\tt glfdiff()}, {\tt fourier\_diffint()} and FIT respectively.}
\label{fig:fo_deriv_func_comp}
\end{figure}

\begin{figure}[h]
\centering
\includegraphics[width=0.75\textwidth]{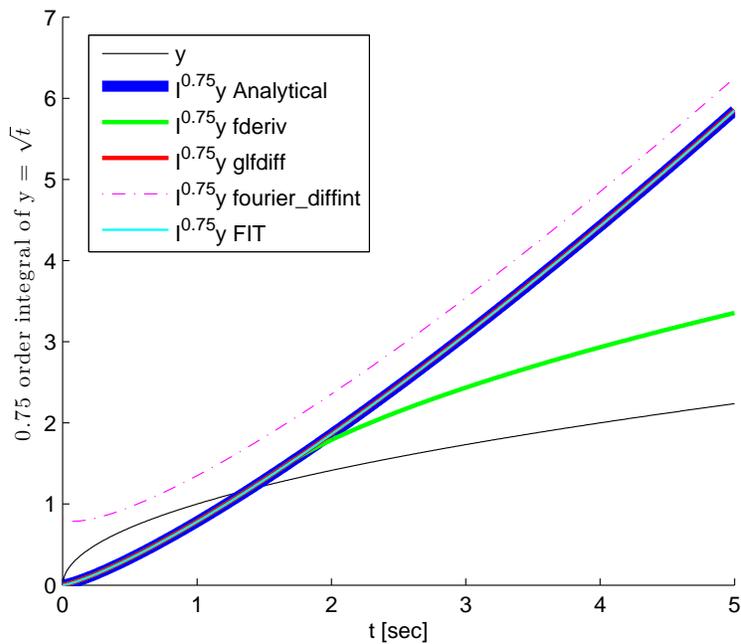}
\caption[Comparison of $0.75^{th}$ order integral of function $y(t) = \sqrt{t}$]{Comparison of $0.75^{th}$ order integration of function $y(t) = \sqrt{t}$, using analytical solution, {\tt fderiv()}, {\tt glfdiff()}, {\tt fourier\_diffint()} and FIT respectively.}
\label{fig:fo_int_func_comp}
\end{figure}

The time steps are all set to 0.01 sec. It can be seen that {\tt fourier\_diffint()} performs not as well as other methods although a big number of Fourier and Gaussian coefficients have been assigned (default values are 260 and 520 for identity polynomial). Its performance on a $3^{rd}$ order polynomial is better. The results are plotted in figures \ref{fig:fo_deriv_func_comp} and \ref{fig:fo_int_func_comp}. Quantitative comparison including computational error and averaged elapsed time (for 20 runs each) are listed in table \ref{tb:func_int_diff_comp}, for the above two problems respectively.

\subsection{Comparison III}
\begin{figure}[h]
\centering
\includegraphics[width=0.8\textwidth]{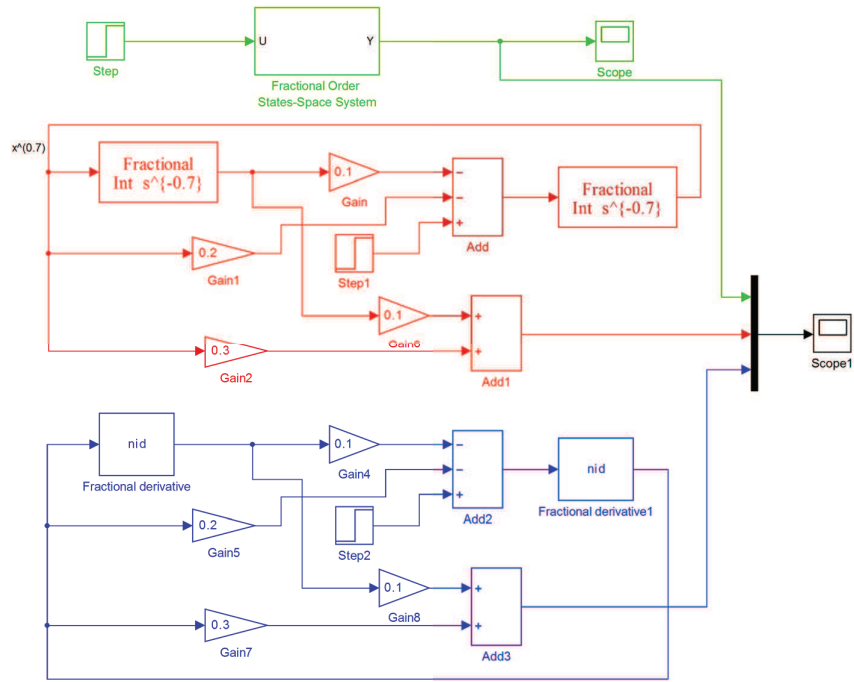}
\caption{The Simulink block diagrams for simulating the FO pseudo state space model in equation (\ref{eqn:FO_SS}).}
\label{fig:Simulink_SS}
\end{figure}

\begin{figure}[h]
\centering
\includegraphics[width=0.75\textwidth]{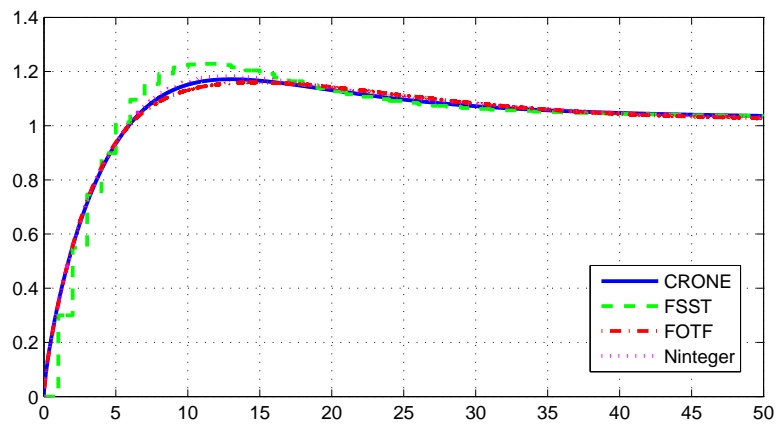}
\caption{Comparison of the simulation results of the FO SS model obtained with different toolboxes.}
\label{fig:FO_SS_comp}
\end{figure}

Although the simulation of fractional order pseudo state space (SS) models can be achieved indirectly, some toolboxes do provide the direct simulation capability, such as the CRONE toolbox and FSST. Since the function {\tt frac\_ss} in CRONE toolbox only adopts the input of commensurate order systems, for comparison purposes, the following commensurate order pseudo state space model is selected,
\begin{eqnarray}
\label{eqn:FO_SS}
\begin{array}{l}
{\left[ {\begin{array}{*{20}{c}}
{{x_1}}\\
{{x_2}}
\end{array}} \right]^{\left( {0.7} \right)}} = \left[ {\begin{array}{*{20}{c}}
0&1\\
{ - 0.1}&{ - 0.2}
\end{array}} \right]\left[ {\begin{array}{*{20}{c}}
{{x_1}}\\
{{x_2}}
\end{array}} \right] + \left[ {\begin{array}{*{20}{c}}
0\\
1
\end{array}} \right]u\\
y = \left[ {\begin{array}{*{20}{c}}
{0.1}&{0.3}
\end{array}} \right]\left[ {\begin{array}{*{20}{c}}
{{x_1}}\\
{{x_2}}
\end{array}} \right]
\end{array}
\end{eqnarray}
To involve more tools into comparison, the FO integrator blocks in the FOTF and Ninteger toolboxes are used to represent the above fractional differential equations in Simulink, as shown in figure \ref{fig:Simulink_SS}. The comparison of the unit step responses computed by the four toolboxes are plotted in figure \ref{fig:FO_SS_comp}, from which it can be seen that the result obtained using FSST ($1$ sec for step size) has bigger difference from the others. However, since analytical solution is not easy to obtain, it is insufficient to claim which method gives highest accuracy. Hence, quantitative comparison is not provided. As an alternative, users can transform the above FO SS model to an FO transfer function model, assuming zero initial conditions,
\begin{eqnarray}
G(s) = C(s^{\alpha}I-A)^{-1}B = \frac{3s^{0.7}+1}{10s^{1.4}+2s^{0.7}+1}.
\end{eqnarray}
Thus, the NILT scripts can be used to compute the numerical solution, which has relatively higher reliability according to the authors observation.

\section{Comments for selection}
\label{sec:com_for_sel}
A tricky part for the simulation of fractional order systems is that even if the system is broken down to the bottom layer, i.e. the analytical solution, it usually still involves the computation of M-L functions, which still needs to rely on the numerical tools or scripts.
From the comparison, it can be seen that in the category of integrating/differentiating a function, glfdiff and FIT outperform other tools in terms of accuracy; in the category of control system simulation, NILT always provides higher accuracy. However, other toolboxes has advantages, for example, ninteger and CRONE toolbox provide integrator blocks in Simulink, which makes the simulation of nonlinear systems possible.

\section{Conclusion}
In this paper, a comprehensive review of the Matlab based numerical tools for fractional calculus and fractional order controls is presented. Quantitative evaluation of the selected tools is conducted. The summarized description and numerical comparison are designated to serve as a reference and guidance for readers when selecting tools for specific applications.

\bibliography{Num_tool_bibfile}

\newpage

\begin{landscape}
\begin{table}[h]
\vspace{0.5 cm}
\caption[Matlab based numerical tools for FC and FO controls]{Matlab based numerical tools for computation of fractional operations and fractional order controls.}
\label{tb:fc_tools}
\begin{center}
\begin{tabular}{|c| l|c|l|c|c|c|c|}
\hline
\#  & Name            & Typical usage               & Sample syntax                 & Author(s)                     & Source            &  Delay     &  MIMO    \\
\hline
\hline
1   & fotf            & FO control toolbox      & {\tt s=fotf('s')}                   & Dingy\"{u} Xue                  & \cite{ref:Xuedingyu_book}  & $\checkmark$ &  Could  \\
\hline
2   & ninteger        & FC and FOC toolbox      & {\footnotesize {\tt nid(k,a,[w1 w2],5,'crone')}}  & D Val\'{e}rio     & \cite{ref:Duarte3}  & Could  & Could\\
\hline
3   & Crone           & FO control toolbox      & {\scriptsize {\tt frac\_tf(1,frac\_poly\_exp(1,0.5) }}& CRONE team    & \cite{ref:CRONE}    & $\times$  & $\checkmark$ \\
\hline
4   & FOMCON          & FO modeling \& control  & {\tt sys\_foss = tf2ss(g)}    & A Tepljakov  & \cite{ref:FOMCON}       & $\checkmark$ &  $\checkmark$   \\
\hline
5a   & mlf            & 2-param M-L func        &  {\tt y=mlf(a,b,-t)}          & I. Podlubny           & \cite{ref:Igor_ML}    &&\\
\cdashline{2-6}
5b   & ml\_func       & $1\sim4$ param M-L func &  {\tt y=ml\_func([a,b,r],-t)} & Dingy\"{u} Xue        & \cite{ref:Monje}           &&\\
\cdashline{2-6}
5c   & ml\_fun        & 2-param M-L func        & {\tt y=ml\_fun(a,b,x,n,e)}    & {\scriptsize S. Mukhopadhyay}   & \cite{ref:Shayok_ml_fun}   &  N/A & N/A\\
\cdashline{2-6}
5d   & gml\_fun       & Generalized M-L func    & {\tt gml\_fun(a,b,r,x,eps0)}  & YQ Chen                 &  \cite{ref:Chen_gml}             &&\\
\cdashline{2-6}
5e   & ml             & 1,2,3-param M-L func    & {\tt e= ML(x,a,b,r)}          & R Garrappa              &  \cite{ref:ML_func}              &&\\
\hline
6a   & NILT           &  Num Inverse of Laplace & Script based                  & L Bran$\rm{\breve{c}\acute{i}}$k  & \cite{ref:Lubomir_NILT_improved}   & $\checkmark$ &  $\times$  \\
\cdashline{2-8}
6b   & INVLAP         &  Num Inverse of Laplace & {\footnotesize {\tt[t,y]=INVLAP('1/s',1,10,100)}}  & Code by Juraj  & \cite{ref:INVLAP_2011_code}& $\checkmark$ & $\times$ \\
\hline
7   & dfod1,2,3       &  Digital FO diff/int    &  {\tt sysdfod=dfod3(n,T,r)}     & I Petr\'{a}\v{s}& \cite{ref:Petras_dfod1}   & N/A & N/A     \\
\hline
8   & irid\_fod ...   &  Impulse Resp Invariant & {\tt df=irid\_fod(-.5,.1,5)}    & YQ Chen   & \cite{ref:Chen_irid_fod}  & N/A &  N/A    \\
\hline
9   & ora\_foc        & Oustaloup-Rec-Approx    &  {\tt ora\_foc(0.5,2,0.1,100)}  & YQ Chen   & \cite{ref:Chen_ora}       & N/A & N/A     \\
\hline
10   & fderiv          & FO diff of r(t)        & {\tt y=fderiv(0.5,r,Ts)}       & F. M. bayat     &  \cite{ref:Fractional_differentiator} & N/A & N/A     \\
\hline
11   & glfdiff         & Finite Diff of G-L     & {\tt y1=glfdiff(y,t,r)}         & Dingy\"{u}  Xue & \cite{ref:Xuedingyu_book}      & N/A & N/A     \\
\hline
12   & {\footnotesize fourier\_diffint} & FO diff of \@f(x) & {\tt \footnotesize fourier\_diffint(f,x,..)}& {\scriptsize G Papazafeiropoulos} & \cite{ref:FO_diff_and_int}   &N/A & N/A \\
\hline
13   & FIT             & FO integration toolbox & {\tt \small fracIntegrationSIM(...)} & Marinov \emph{et al.}         & \cite{ref:FIT_download}       & N/A & N/A  \\
\hline
14   & DFOC            & Discrete FO PID        &  {\tt \small DFOC(K,Ti,Td,m,d,Ts,n)} & I Petr\'{a}\v{s}   & \cite{ref:Petras_DFOC} & N/A & $\times$ \\
\hline
15   & FOPID           &  FO PID                &  ---                          & Lachhab \emph{et al.}         & \cite{ref:Nabil_FOPID}         & --- & $\times$ \\
\hline
16   & FOCP            &  {\footnotesize Fractional optimal control}   & Calling RIOTS  &{\small C Tricaud \emph{et al.}} & \cite{ref:FOCP_Tricaud} & $\times$ & $\checkmark$ \\
\hline
17   & FSST            &  FO S-S Toolkit        &  Simulink blocks              & D. Sierociuk  & \cite{ref:fsst_1_7}   & $\checkmark$ & $\checkmark$ \\
\hline
18   & FVO             & {\footnotesize F}ractional variable order  & {\tt ban(alpha,N,h)}   & Podlubny \emph{et al.} & \cite{ref:Igor_matrix_approach}   & N/A & N/A     \\
\hline
19   & forlocus        & RL plot of FO TFs      & {\tt forlocus(num,den,l)}   & Zhuo Li \emph{et al.}        & \cite{ref:FO_RL_mine}                             & N/A & N/A     \\
\hline
\multicolumn{8}{l}{The `Delay' column denotes if the script/toolbox is able to handle time delay in the FO model.} \\
\multicolumn{8}{l}{The `MIMO' column denotes if the script/toolbox is able to handle MIMO FO models.}
\end{tabular}
\end{center}
\end{table}
\end{landscape}


\clearpage

\begin{table}[h!]
\caption{Quantitative evaluation results for the test problems 1$\sim$5.}
\label{tb:tools_eval_error}
\small
\begin{center}
\begin{tabular}{c|l|l|l|l|l}
\hline
\backslashbox[-5pt][l]{Method \kern-3em}{\kern-1em Error}
    &1                      &2              &3          &4           &   5   \\
\hline
M   &0                      &-              &-          & -          &   -   \\
\hline
1   &$1.4955$                  & $8.4176$   &  $6.6813$ & ``0''          &   ``0''   \\
\hline
2   &$3.18\!\!\times\!\!10^{-13}$  & $2.5287$ &$0.3831$ &  $9.8434$  &   $2.5519$    \\
\hline
3   &$0.4956$               & $2.9627$      & $1.4254$  &  $10.454$   &   $3.1857$ \\
\hline
5a  & 0                     &  -            & $4.69\!\!\times\!\!10^{-4}$   &  -          &   -           \\
\cdashline{1-6}
5c  & $8.62\!\!\times\!\!10^{-12}$  &  -    & $1.08\!\!\times\!\!10^{-10}$         &  -          &   -           \\
\hline
6a  & $0.0016$                  &  $0.0236$ & $ 0.0206$ &  $6.1528$   &  $2.4477$     \\
\cdashline{1-6}
6b  & $0.0059$                  &  $0.0012$ & $2.49\!\!\times\!\!10^{-5}$  &  $3.4722$   &  $1.5042$     \\
\hline
8   & $0.5327$                  &  $0.0071$ & $0.2189$  &  -          &  -            \\
\hline
\end{tabular}
\end{center}
\end{table}

\begin{table}[h]
\caption{Quantitative comparison of function int/diff tools.}
\label{tb:func_int_diff_comp}
\small
\begin{center}
\begin{tabular}{r|c|c|c|c|c}
\hline
\backslashbox[-5pt][l]{Criteria \kern-3em}{\kern-1em Methods}
    &  Analytical       &{\tt fderiv()}&   {\tt glfdiff()} & {\tt fourier\_diffint()} & FIT \\
\hline
Error 1   &-              &140.5000   & 1.8232    &  792.4660 & 0.0000  \\
\hline
Elapsed T1&0.0001         &1.4028     &0.0029     &  0.0874   & 0.0209  \\
\hline
\hline
Error 2   &-              &339.7973   &2.1208     &  250.5433 & 0.0743  \\
\hline
Elapsed T2&0.0001         &1.4105     &0.0029     &  0.0893   & 0.0201  \\
\hline
\end{tabular}
\end{center}
\end{table}

\end{document}